\documentclass[conference]{IEEEtran}
\IEEEoverridecommandlockouts
\usepackage{hyperref}

\usepackage{cite}
\usepackage{amsmath,amssymb,amsfonts}
\usepackage{algorithmic}
\usepackage{graphicx}
\usepackage{textcomp}
\usepackage{xcolor}
\def\BibTeX{{\rm B\kern-.05em{\sc i\kern-.025em b}\kern-.08em
    T\kern-.1667em\lower.7ex\hbox{E}\kern-.125emX}}
\begin{document}

\title{A Transformer-based Model to Detect Phishing URLs}

\author{\IEEEauthorblockN{ Pingfan Xu}
\IEEEauthorblockA{\textit{School of Computer Science} \\
\textit{University of Guelph}\\
Guelph, Canada \\
pingfan@uoguelph.ca}

}

\maketitle

\begin{abstract}
Phishing attacks are among emerging security issues that recently draws significant attention in the cyber security community. There are numerous existing approaches for phishing URL detection. However, malicious URL detection is still a research hotspot because attackers can bypass newly introduced detection mechanisms by changing their tactics. This paper will introduce a transformer-based malicious URL detection model, which has significant accuracy and outperforms current detection methods. We conduct experiments and compare them with six existing classical detection models. Experiments demonstrate that our transformer-based model is the best performing model from all perspectives among the seven models and achieves 97.3\% of detection accuracy.
\end{abstract}

\begin{IEEEkeywords}
Transformer, Phishing URL Detection, Machine Learning, Deep Learning
\end{IEEEkeywords}

\section{Introduction}
 Phishing is a type of attack leveraged by cyber criminals which impersonate some legitimate organizations' websites or URL to deceive the victims \cite{RzeszutBachrach}. The attackers will use somebody or some company's identity that the people trust to send them spam emails or messages embedded with those phishing links \cite{JakobssonMyers}. The count of phishing-related cyber crimes incidents proliferation in the last several decade. It was mentioned in Verizon Data Breach Investigation Report (DBIR) that phishing attacks caused more than 22\% of the breaches in 2020 \cite{dbir}. The consequent ransomware was estimated to cost over 8 billion USD all over the world \cite{ShahbazenezhadKolini}. Among the 22\% of the breaches in 2020 caused by phishing attack, 99\% of them were delivered through emails or websites. In December 2013, 42890 unique phishing websites were recorded and reported worldwide \cite{apwg2017phishing}. Since phishing emails and website both may involve malicious URLs, the exploration of malicious URL detection has double benefits for both anti-email-phishing and anti-website-phishing. 
 
In order to mitigate phishing attacks, there are already numerous anti-phishing solutions in practice. As for modern web browsers, there are indicators as built-in countermeasures for a phishing website. For instance, browsers would validate the website's SSL certificate and provide the result by displaying a special indicator icon on the address bar for users' references. Unfortunately, this validation process could be easily bypassed if the attacker used an URL visually similar to the original site's address. Studies showed that these indicators might be ineffective and even put the users into a higher-risk situation \cite{whalen2005gathering, lin2011does, egelman2009trust}. Another commonly employed method against phishing is two-factor authentication. It introduces an extra security layer. One of the widely used techniques for two-factor authentication (2FA) is sending real-time generated verification code via SMS \cite{ApandiSallim}. Even though 2FA may protect against phishing attacks, it typically requires a different device. Additionally, there are researches found possible ways to bypass 2FA as well \cite{DmitrienkoLiebchen}. Therefore, it would be critical to discover reliable and lightweight malicious URLs detection solutions to identify malicious URLs on time to prevent severe consequences and potential losses. 
 
 This paper's contributions are as following:
 \begin{itemize}
     \item We reviewed and categorized the typical existing approaches for phishing URL detection.
     \item We proposed an innovative transformer-based model as a light-weight but high performance solution for malicious URL prediction.
     \item We evaluated our proposed model by comparing its performance with six other existing classical models'. We proved that our model has better performance competing with these other models.
 \end{itemize}
 In this paper, Section \ref{sec:bg} provides a background introduction and literature reviews of the existing anti-phishing solutions in practice. Then, our proposed solution and its design details will be illustrated at Section \ref{sec:proposedmodel}. Afterwards Section \ref{sec:evaluation} will explain how we train our Transformer model. The experiment section will be followed by the model evaluation section where we mainly compare the performance of six machine learning models (Decision Tree, Random Forest, Multi-layer Perceptrons, XGBoost, Support Vector Machine, Auto Encoder) implemented by Sundari and her team \cite{Sundari} to our transformer model. Finally, we will provide some suggestions to improve our proposed solution further at Section \ref{sec:conclusion}.

\section{Literature Review and Background}\label{sec:bg}
 In recent years, researchers have explored a massive amount of possible solutions for phishing website detection. We roughly categorized the existing detection techniques into content-based, property-based, and URL-based approaches targeting different websites. Each approach has its advantages and disadvantages. Several typical pieces of research under each approach category will be reviewed in this section. Additionally, as one innovative solution under URL-based approaches, the transformer model will be demonstrated with its background.
 
 \subsection{Literature Review}
 Content-based phishing website detection may be one of the most intuitive techniques among the three. It might also be the method closest to how the average person gets to determine phishing sites. Firstly, people would visually identify the identity that the images and texts on the website suggested. With the identified site identity, they could query the corresponding URL of the identity through a trusted source. The website would be considered a benign one only when the queried URL matches the URL shown on the address bar. The logic behind content-based phishing website detection techniques was the same. For example, Chiew et al. utilized the website logo for phishing site detection \cite{ChiewChang}. The researchers employed Google image search on extracted logos from the website to get the portrayed site identity. The phishing detection result was determined through the consistency check of the portrayed site identity and the actual domain of the testing website. Their proposed solution provided an outstanding performance with 99.8\% true positive rate and 87.0\% true negative rate.
Additionally, the texts shown on a site could also be used for phishing detection. Yang et al. proposed a multidimensional feature-based detection technique that included vectorizing texts on the page for determine the possibility that the vectorized text was from a phishing site \cite{YangZhao}. Their approach achieved 98.99\% in its accuracy. Similarly, there was another approach using the same methodology and the website favicon for phishing detection \cite{ChiewChoo} where its true positive rate reached 96.93\%.
 
 In addition to visual and literal contents on a web page, other properties or meta-data information hidden to everyday users could also be employed for phishing detection. The property-based approach generally pays attention to the abnormal behaviors of a website. For example, Pan and Ding used six abnormal properties, including Abnormal URL, Abnormal DNS record, Abnormal Anchors, Server-Form-Handler, Abnormal cookie, and Abnormal Secure Sockets Layer(SSL)-certificate, to conduct phishing checking \cite{PanDing}. Their proposed model gives an average false positive rate as low as 4\%. Since studies suggested the number of redirect pages contained in malicious sites is between 2 to 4 \cite{MohammadThabtah}, the count of redirecting times was proposed as one possible detecting approach. To avoid being easily identified by end-users inspecting source code, attackers may use JavaScript to disable the mouse right-click function \cite{Harinahalli}. Hence, uncommon disabling of browser functionalities could be another standard for detection. Furthermore, other website properties related to its domain information can be utilized. Properties like the domain age, domain name, and IP address were also proven to have positive contributions to phishing detection in recent researches \cite{FetteSadeh, ZamirKhan}. 
 
 Besides content-based and property-based approaches, the URL-based method is also a hot research topic in the field of phishing detection. URL-based approaches take advantage of URL analysis to conduct phishing predictions. Researchers have tried different URL-based techniques to detect malicious URLs. Most of them utilize hand-crafted features extracted from the URLs. Ma et al. proposed a detection method from the lexical and host-based features in 2011, which detects malicious Web sites from a balanced dataset with 99\% accuracy \cite{Ma}. Choi et al. identified malicious URLs with six manually extracted class features from URLs: lexical, link popularity, webpage content, DNS, DNS fluxiness, and Network \cite{Choi}. Their solution was capable of malicious URL detection with over 98\%  of accuracy. In 2012, Zhang and Wang developed an accurate, real-time, and language-independent malicious URL classifier with hosted and lexical features of target URLs \cite{Zhang}. A model based on word n-gram and Markov Chains that could generate proactive blacklists of malicious URLs was designed by Marchal et al. in 2012 \cite{Marchal}. Their approach to preventing phishing scams could be considered the opposite of traditional detections. Pao et al. proposed a detection method that utilizes the estimation of conditional Kolmogorov complexity of URLs \cite{Pao}. Their approach could independently handle large amounts of input URLs very efficiently with high accuracy. Additionally, their model could be used in conjunction with other URL-feature-based detection techniques to improve the prediction accuracy further. 
 
 Since URLs are essentially pieces of texts, researchers started exploring the possibilities of utilizing Natural Language Processing (NLP) techniques for URL-based phishing detection. As one of the most innovative NLP models in recent years, the transformer model has also been applied to phishing site detection. Researches applied transformer model on malicious URL detection turn out to have outstanding accuracy. It was proven that the novel transformer model would have comparably good performance to the more traditional approaches \cite{RuddAbdallah}.
 
 \subsection{Background}
 In recent years, Transformer is a hot research topic in handling tasks closely related to characters. One of the most crucial concepts in the transformer model is called attention. In 2017, the Google machine translation team published a research paper, ``Attention Is All You Need." \cite{Vaswani} It completely subverted some traditional network structures such as RNN and CNN and only applied the attention mechanism to perform machine translation tasks. The transformer model had a better performance than other machine learning models, and this let the attention technique became a research hotspot. Later, Devlin et al. introduced an improved language processing model, Bidirectional Encoder Representations from Transformers (BERT) \cite{Jacob}, based on the Transformer. It is different from the standard language representative models because BERT is used to pre-train deep bidirectional representations to produce an encoder with hidden states that can be used for various NLP fine-tuning tasks. In 2018, Radford et al. from OpenAI came up with the Generative Pre-trained Transformer (GPT) series models \cite{Radford}. Their approach is similar to BERT but utilizes L - R masked modeling. The decoder stack performs a prediction on the following elements of the sequence from the previous token.
 
 \begin{figure}[htbp]
 \centerline{\includegraphics[width=3.5in]{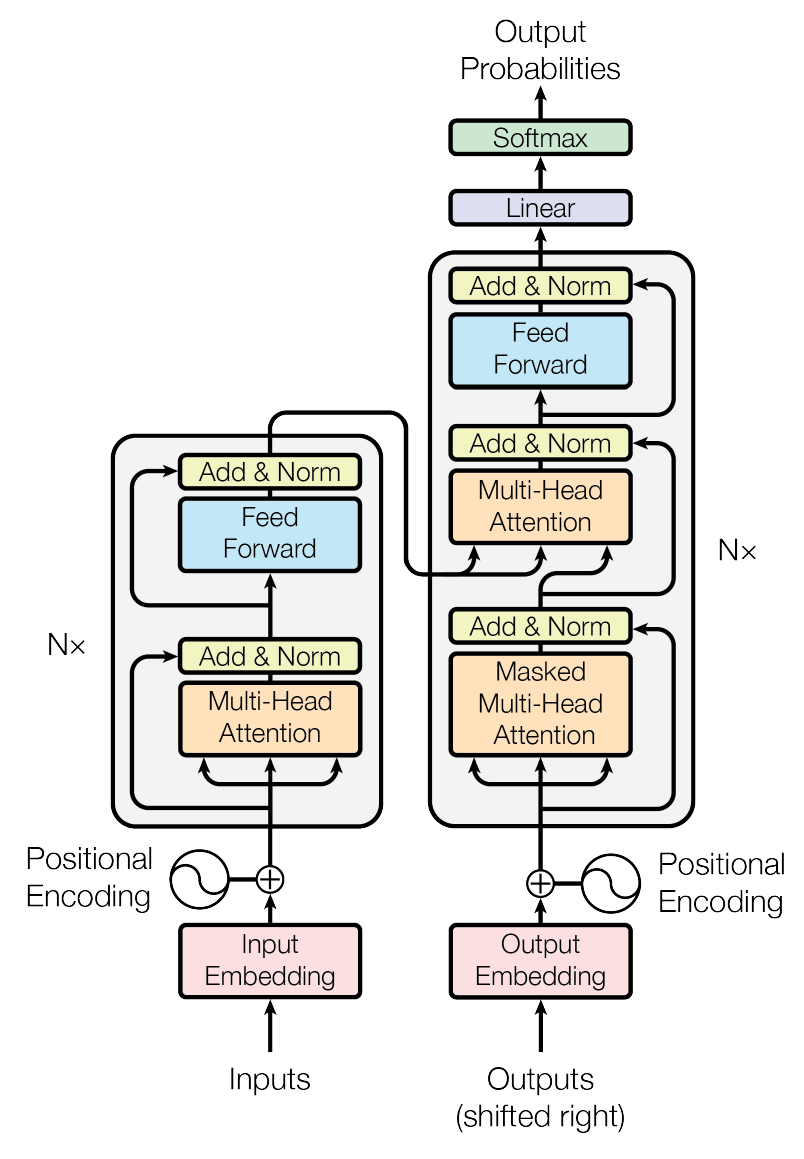}}
 \caption{Standard structure of Transformer.}
 \label{fig1}
 \end{figure}
 
\begin{figure}[htbp]
 \centerline{\includegraphics[width=1.5in]{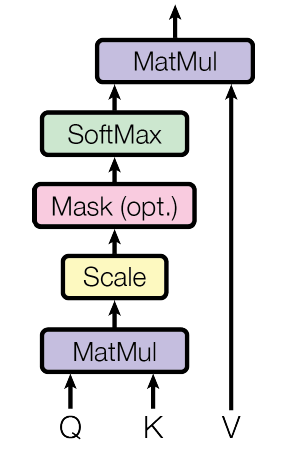}}
 \caption{Self-attention mechanism.}
 \label{fig2}
 \end{figure}

 A typical transformer model contains two stacks: the encoder and the decoder. Each stack of it includes an embedding layer, positional encoding, multi-headed attention layers, feed-forward layers, residual connections, and masks (as depicted in Fig. \ref{fig1}). The encoder takes the input text and makes it model-readable through input embedding and positional encoding. The encoder then applied several tensors to the processed input. The tensors are made of multi-head attention layers connected to feed-forward layers, which provides the self-attention feature to the model as a combination. The multi-head attention layers are the collection of identical self-attention mechanisms. Each self-attention mechanism (as depicted in Fig. \ref{fig2}) can be described as a mapping:
 \begin{equation}
 Attention(Q,K,V)=softmax(\frac{QK^{T}}{\sqrt{d_{k}}})\label{eq1}
 \end{equation}
 where query (Q), key (K), value(V) and output are all matrices and $d_{k}$ is the dimension of input queries and keys. 
 
\begin{figure}[htbp]
 \centerline{\includegraphics[width=2in]{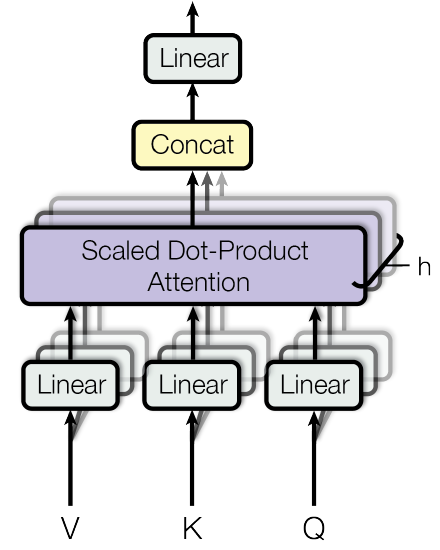}}
 \caption{Multi-head attention.}
 \label{fig3}
 \end{figure}

 The multi-head attention (as depicted in Fig. \ref{fig3}) can then be described as:
 \begin{equation}
 MultiHead(Q,K,V)=Concat(head_{1},...,head_{h})W^{O}\label{eq2}
 \end{equation}
 where \\\\
 $head_{i}=Attention(Q{W_{i}}^{Q},K{W_{i}}^{K},V{W_{i}}^{V})$,\\ ${W_{i}}^{Q}\in\mathbb{R}^{d_{model}\times d_{k}}$,\\ ${W_{i}}^{K}\in\mathbb{R}^{d_{model}\times d_{k}}$,\\ ${W_{i}}^{V}\in\mathbb{R}^{d_{model}\times d_{v}}$,\\ $W^{O}\in\mathbb{R}^{{hd_{v}}\times d_{model}}$. \\
 
 Then, the result from multi-head attention is passed to a fully connected feed-forward network (FFN) containing two linear transformations with a Rectified Linear Unit (ReLU) as activation function in between:
 \begin{equation}
 FFN(x)=ReLU(W_{1}x+b_{1})W_{2}+b_{2}\label{eq3}
 \end{equation}
 The output of FFN is considered the encoder's output and is passed to the decoder for an intermediate tensor operation. The decoder has a similar structure as the encoder. The primary differences are the masking operation and an additional multi-head attention layer on each tensor. The masking operation is applied on the first multi-head attention layer, which takes the processed input of the decoder. The additional multi-head attention layer is the one next to the masked multi-head attention layer. Also, the extra layer is where the output of the encoder is taken into the decoding operation.
 
 The above usage based on the network architecture of the transformer is most frequently applied to NLP problems. It can be seen that the transformer model has certain advantages compared with other models in processing language semantics. However, using the transformer in the information security area, especially the prediction and classification of URLs, is a relatively new attempt. Rudd and Abdallah from FireEye Inc. came up with the idea of using a transformer model for malicious URL prediction \cite{RuddAbdallah}. They split each URL into single-character tokens and append a classification token ``CLS" at the end. To solve the different lengths of each URL, padding tokens ``PAD" will be added to normalize the input to a fixed length for training. The preprocessed URLs will be projected into an embedding area with a stack of attention layers and feed-forward neural network layers. Like the transformer's basic idea, each input token has an essential relation with other tokens no matter how long the distance is.
Meanwhile, the feed-forward layers allow learning the combination of the input tokens' relationship and their context. The training process utilized over a million labeled malicious and benign URLs. For the evaluation, they compared their transformer model with four other models: Random Forest, CNN, and LSTM. Finally, they concluded that the transformer model had the highest accuracy score. Our model that will be later introduced in this paper is inspired by their design.

 \section{Proposed Model}\label{sec:proposedmodel}
 Based on the existing solutions and approaches of phishing or malicious URL detection, our group has proposed and implemented an innovative solution mainly based on the transformer model. The design and implementation details of our approach will be introduced in this section.
 
 Our approach to the transformer model design is not identical to the standard structure. The model design of Rudd and Abdallah from FireEye Inc. inspires the transformer model design of our approach \cite{RuddAbdallah}. In our solution, the transformer model is very similar to the design of OpenAI's GPT model, one of the famous variants of the Transformer model. Our Transformer model applied left to right (L-R) modeling and only contained the encoder part from the standard transformer model. There are mainly two parts included in our solution: the input text pre-processing part and the classifier model part.
 
 Specifically for the input text pre-processing part, we tokenize the input URL first by splitting it into single-character tokens. The most frequently appeared single-character tokens in the URLs of the training dataset are used to form the token repository, which is also known as the vocabulary of the tokenizer. The maximum vocabulary size is designed to be 256 is corresponding to a 256-character embedding space from index 0 to 255. Any characters contained in the input URL out of the most frequently appeared 256 characters will be replaced by a unique character $<OOV>$. For the relatively shorter input URLs, they will be padded to the max length of 256. On the other hand, the input URLs that are longer than 256 characters will be truncated to the length of 256. Therefore, after the input text pre-processing operations, each input URL will be tokenized to an array of length 256. For example, given an input URL \\\\$https://www.google.com/$\\\\ the tokenized array will be\footnote{\textit{Note: A character's corresponding sequence number representation may differ depending on the character's appearance frequency in the given dataset. Furthermore, $0$ is the padding's corresponding sequence number representation.}}: \\\\$[15,  2,  2, 13,  9, 31,  4,  4, 33, 33, 33, 18, 26,  5,  5, 26, 17,\\
        3, 18, 12,  5, 16,  4,  0,  0,  0,  0,  0,  0,  0,  0,  0,  0,  0,
        0,  0,  0,  0,  0,  0,  0,  \\0,  0,  0,  0,  0,  0,  0,  0,  0,  0,
        0,  0,  0,  0,  0,  0,  0,  0,  0,  0,  0,  0,  0,  0,  0,  0,  \\0,
        0,  0,  0,  0,  0,  0,  0,  0,  0,  0,  0,  0,  0,  0,  0,  0,  0,
        0,  0,  0,  0,  0,  0,  0,  0,  \\0,  0,  0,  0,  0,  0,  0,  0,  0,
        0,  0,  0,  0,  0,  0,  0,  0,  0,  0,  0,  0,  0,  0,  0,  0,  0,
        \\0,  0,  0,  0,  0,  0,  0,  0,  0,  0,  0,  0,  0,  0,  0,  0,  0,
        0,  0,  0,  0,  0,  0,  0,  0,  0,  \\0,  0,  0,  0,  0,  0,  0,  0,
        0,  0,  0,  0,  0,  0,  0,  0,  0,  0,  0,  0,  0,  0,  0,  0,  0,
        0,  \\0,  0,  0,  0,  0,  0,  0,  0,  0,  0,  0,  0,  0,  0,  0,  0,
        0,  0,  0,  0,  0,  0,  0,  0,  0,  0,  \\0,  0,  0,  0,  0,  0,  0,
        0,  0,  0,  0,  0,  0,  0,  0,  0,  0,  0,  0,  0,  0,  0,  0,  0,
        0,  0,  \\0,  0,  0,  0,  0,  0,  0,  0,  0,  0,  0,  0,  0,  0,  0,
        0,  0,  0,  0,  0,  0,  0,  0,  0,  0,  0,  \\0,  0,  0,  0,  0,  0,
        0]$\\\\

For the classifier model part of our Transformer model, there are eight main layers connected sequentially. The first layer is the input layer that takes the pre-processed input URL as an array 256. The input layer passes the tokenized input URL to a token and position embedding layer. The dimension of this layer is equal to the vocabulary size of the dataset gotten after the input text pre-processing operations. Thus, the dimension of this layer is up to 256. This token and position embedding layer consists of two sub-layers which are both embedding layers. One of these two separate embedding layers is for tokens, while the other is for token index (positions). The layer next to the token and position embedding layer is a transformer block. The transformer block is the same as the encoder layer of the standard transformer model, which consists of sublayers: multi-head attention and point-wise feed-forward networks. The transformer block has the same embedding dimension as the token and position embedding layer. There are four attention heads included in the Transformer block. The relationship of model dimension ($d_{model}$), key dimension ($d_{k}$), and value dimension ($d_{v}$) is: \\\\
$d_{k}=d_{v}=d_{model}/4=64$. \\\\
Thus, the multi-head attention sublayer can be described very similar to (\ref{eq2}) as:
\begin{equation}
\resizebox{\columnwidth}{!}{
$MultiHead(Q,K,V)=Concat(head_{1},head_{2},head_{3},head_{4})W^{O}$
}
\label{eq4}
\end{equation}
where \\\\
$head_{i}=Attention(Q{W_{i}}^{Q},K{W_{i}}^{K},V{W_{i}}^{V})$,\\ ${W_{i}}^{Q}\in\mathbb{R}^{256\times 64}$,\\ 
${W_{i}}^{K}\in\mathbb{R}^{256\times 64}$,\\ 
${W_{i}}^{V}\in\mathbb{R}^{256\times 64}$,\\ $W^{O}\in\mathbb{R}^{4\times64\times256}$.\\

Inside the transformer block, the hidden layer size in the feed-forward network is 128. The transformer layer outputs one vector for each time step of our tokenized input URL. We take the mean across all time steps and use a feed-forward network on top of it to classify text. Also, the dropout rate of 0.1 is employed to help with preventing overfitting. At the end of our classifier model, a softmax function is employed to generate the final prediction result of whether the input URL is malicious or benign.

\section{Experiment}\label{sec:evaluation}

\subsection{Dataset}
Our dataset is a combination of two resources: PhishTank \cite{phishtank}, and the University of New Brunswick (UNB) \cite{unb}. PhishTank provides hourly updated phishing URLs sets in various formats including .json, .csv etc. The phishing URLs on PhishTank are verified as valid by the community, which gives better quality on the original dataset. PhishTank is a widely used source of phishing URLs for researches related to phishing detection \cite{LiewFazlida, BarracloughHossain, VarshneyMisra}. The other dataset from UNB has a collection of benign URLs, spam URLs, phishing URLs, malware URLs, and defacement URLs. Researches like \cite{MamunRathore} used this dataset as part of its input URL feeding source. Since we only focus on phishing and legitimate URLs, we pulled malicious URLs from PhishTank and benign URLs from UNB. We randomly selected 20000 URLs with 10000 benign URLs from UNB's dataset and 10000 malicious from PhishTank's dataset for training and testing purposes. All selected URLs are labeled with $0$ or $1$ for benign or malicious correspondingly. We separate 80\% (16000 URLs) for training and 20\% of the data (4000 URLs) for testing. Before the training process, we also shuffled our labeled dataset to improve our model's robustness.

\subsection{Training Process}
\begin{figure}[htbp]
 \centerline{\includegraphics[width=3.5in]{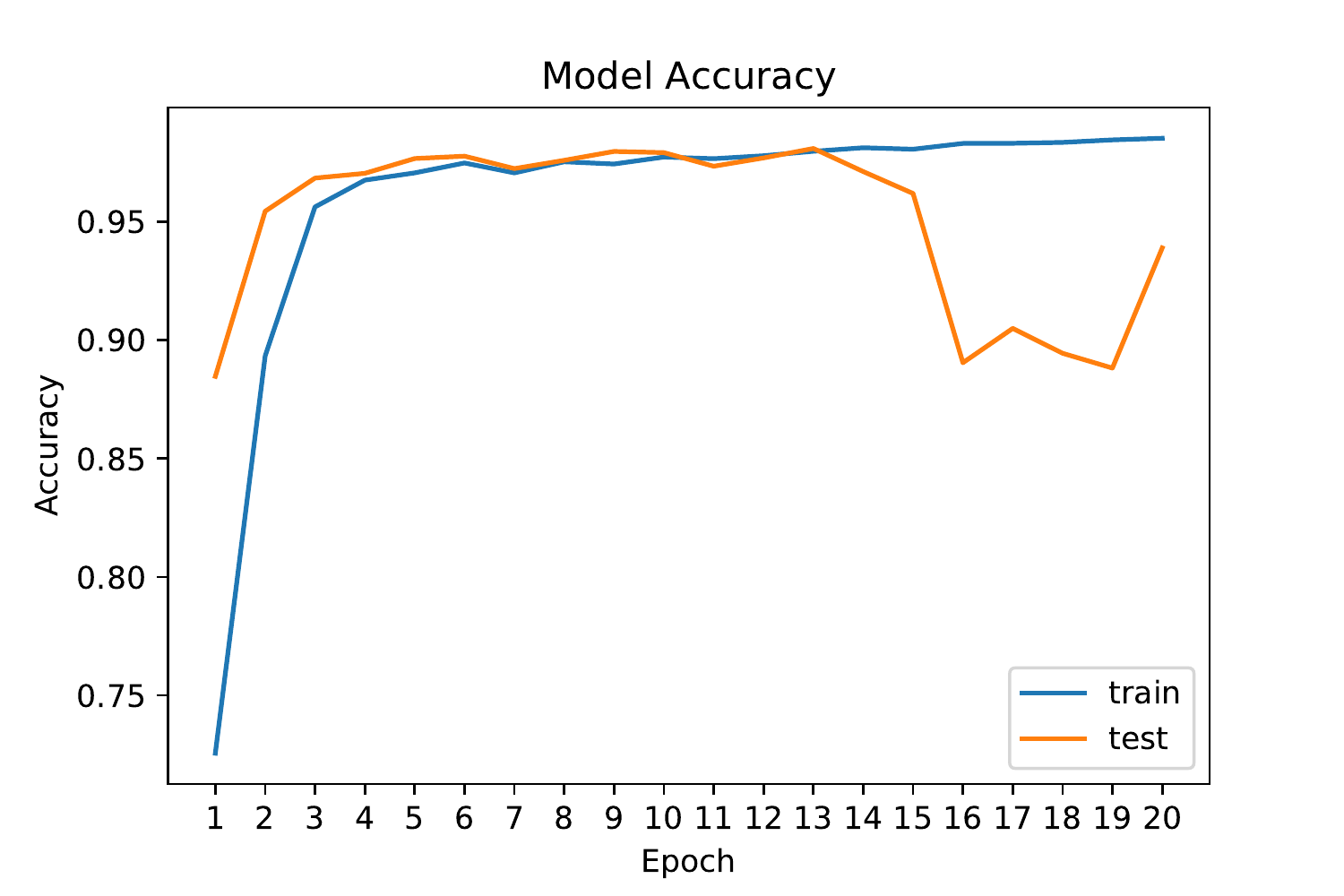}}
 \caption{Transformer model accuracy during training \& validation.}
 \label{fig4}
 \end{figure}

\begin{figure}[htbp]
 \centerline{\includegraphics[width=3.5in]{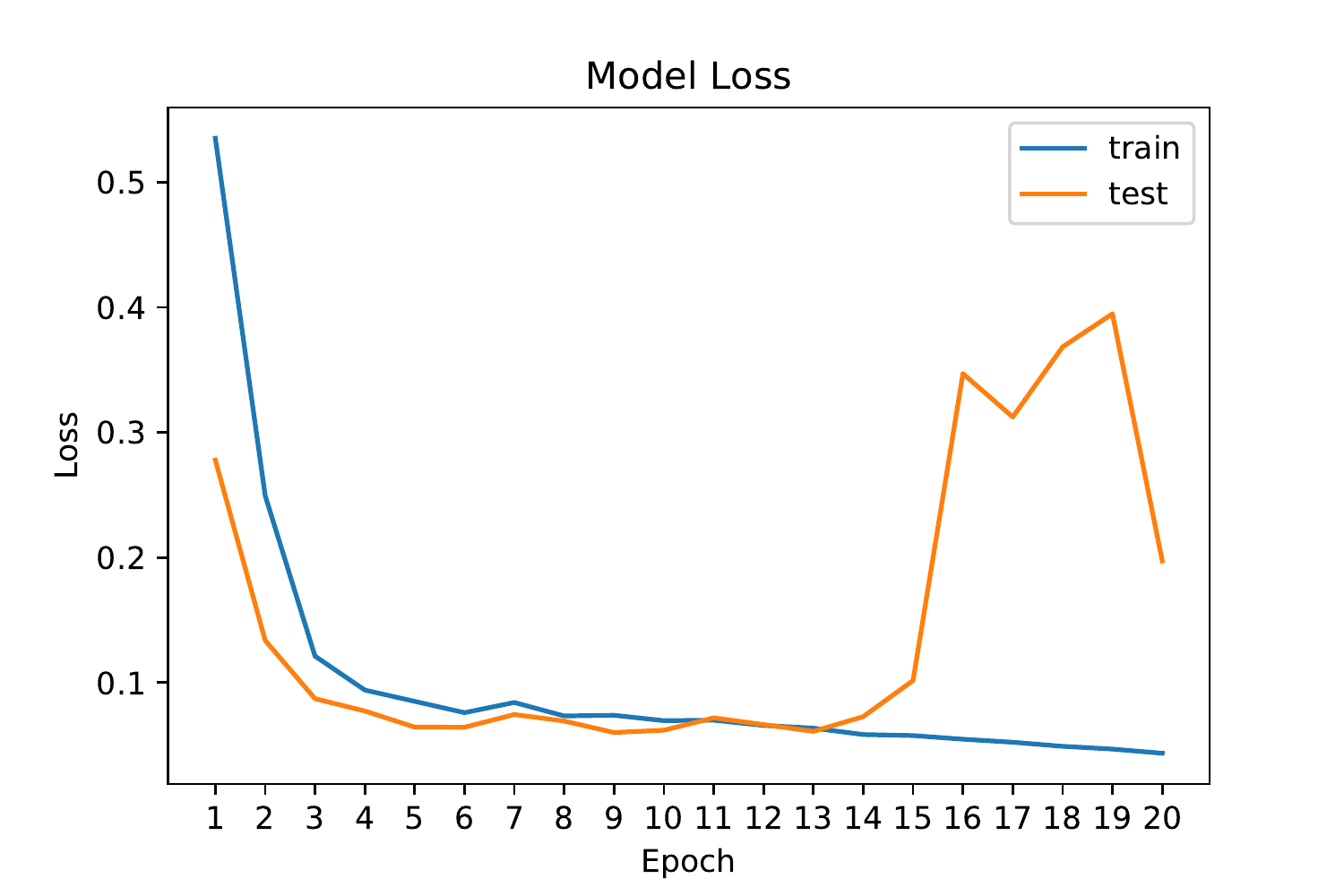}}
 \caption{Transformer model loss during training \& validation.}
 \label{fig5}
 \end{figure}

Initially, we used a batch size of 512 and 20 epochs for test training to get the appropriate epoch number to avoid overfitting. After each training epoch, a checkpoint would be saved for referring back at the model finalization stage. Based on the initial training experiment, we found that both the training and testing accuracy were kept on a high (over 95\%) and steady level between the $3^{rd}$ and $13^{th}$ (as depicted in Fig. \ref{fig4}). Additionally, the validation accuracy of our transformer model started dropping dramatically after the $13^{th}$ epoch of training (as depicted in Fig. \ref{fig4}). At the same time, the transformer model's loss began increasing significantly (as depicted in Fig. \ref{fig5}). According to this finding, it suggested that the checkpoint saved immediately after the $13^{th}$ epoch of training could be considered as a good balance between pursuing high accuracy and avoiding model overfitting. Thus, the checkpoint saved after the $13^{th}$ epoch of training were taken as the finalized training result of our model.
 
 \section{evaluation}
 As mentioned in the previous section, the checkpoint after the $13^{th}$ epoch of training was chosen as the finalized model based on the 20-epoch experimental training results of model accuracy and model loss. Thus, the evaluation analysis in this section would be conducted on top of the transformer model's checkpoint after the $13^{th}$ epoch of training. 
 
 \subsection{Model Evaluation}
\begin{figure}[!t]
\centering
\includegraphics[width=3.45in]{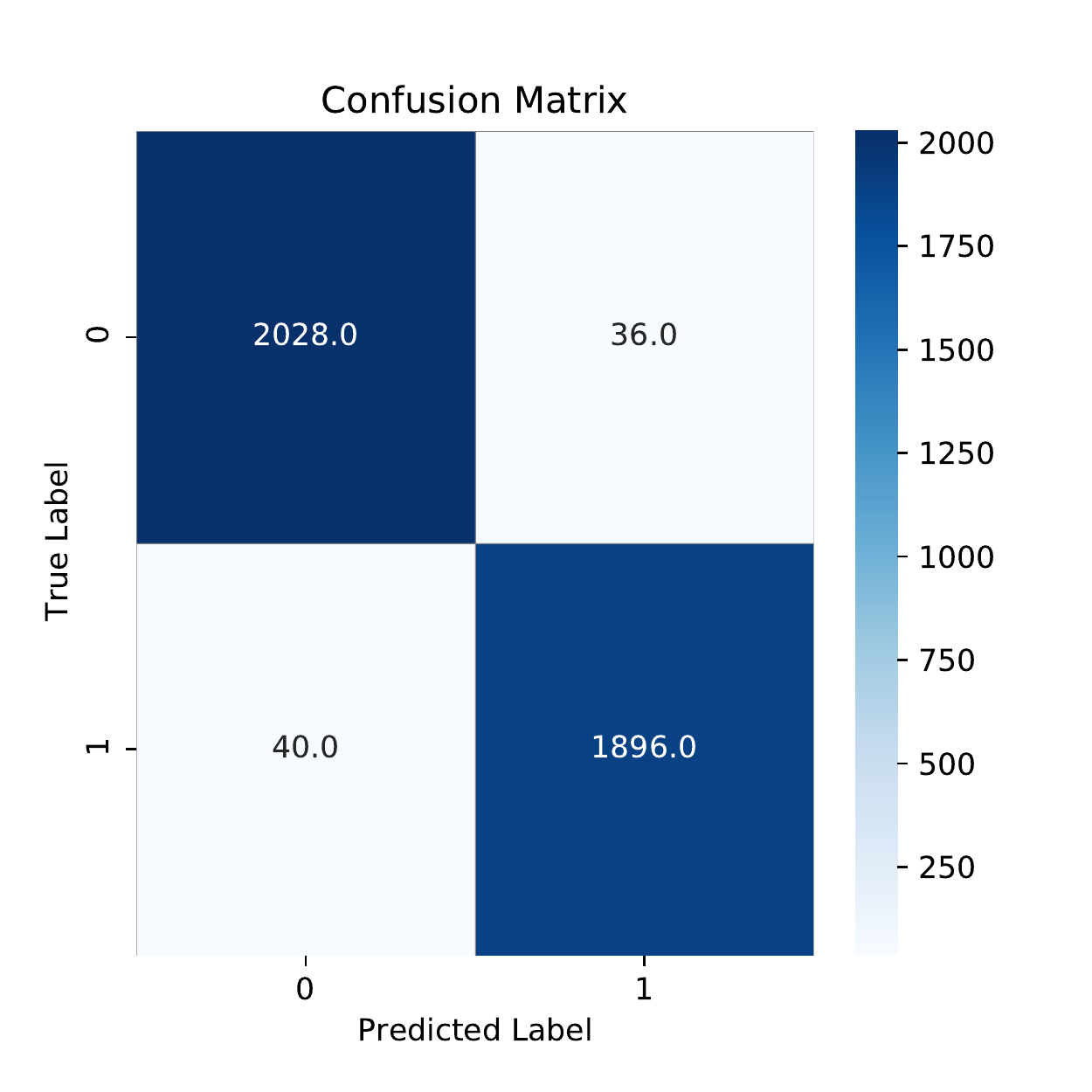}
\caption{Confusion matrix of Transformer model.}
\label{fig6}
\end{figure}

We ran the prediction on the validation dataset that includes 4000 records combined with benign and malicious URLs to evaluate our model. Then, we used the prediction results and the original labels to generate the confusion matrix. In our experiment, True Negative (TN) is the number of benign URLs classified as benign, False Positive (FP) is the number of benign URLs misclassified as malicious, False Negative (FN) is the number of malicious URLs misclassified as benign, and True Positive (TP) is the number of malicious URLs classified as malicious. Fig. \ref{fig6} shows the confusion matrix.

In terms of confusion matrix results, the ideal situation is that it contains a large number of TP and TN and a small number of FP and FN. From Fig. \ref{fig6}, we have a large number of TP (1896). And we also have a high value for TN (2028). These two numbers indicates our model is capable of accurately predict both malicious and benign URLs. Additionally, we can also find that neither FP (36) nor FN (40) is a large number. These two numbers suggest there is higher chance that the input URL is actually benign when our transformer model made predictions incorrectly. Even though our transformer model may make mistakes while predicting, its overall performance is outstanding based on the results that its confusion matrix shows.

To further evaluate our Transformer model, we utilized the following indicators: accuracy (\ref{eq5}), precision (\ref{eq6}), recall (\ref{eq7}), and F1-score (\ref{eq8}). Accuracy is the most intuitive metric that calculates correct predicted results over all the predictions. Precision measures the ratio of precise positive prediction overall positive prediction. Recall shows the correct positive prediction over the value of a specific class. Furthermore, F1-score calculates the weighted average of the precision and recall. According to the confusion matrix, the accuracy is 0.981, the precision is 0.981, the recall is 0.979, and the F1-score is 0.980. These values all together indicate the high-quality performance of our model.

\begin{equation}
\label{eq5}
Accuracy = \frac{TP + TN}{TP + FN + TN + FP}
\end{equation}

\begin{equation}
\label{eq6}
Precision = \frac{TP}{TP + FP}
\end{equation}

\begin{equation}
\label{eq7}
Recall = \frac{TP}{TP + FN}
\end{equation}

\begin{equation}
\label{eq8}
F1-score = 2 * \frac{Precision * Recall}{Precision + Recall}
\end{equation}

\subsection{Model Comparison}
In addition, to evaluate our Transformer model by its validation performance, we also compared our model's performance with other typical classifier models. We employed six extra trained models from Sundari, and her team \cite{Sundari}'s work. Their proposed approach could be considered as a hybrid solution of property-based and URL-based approaches. They mainly extracted 17 features from each URL that belongs to three main categories: Address-Bar-based Features, Domain-based Features, and HTML-\&-Javascript-based Features. Our Transformer model is comparable with their six models because their models were also trained on the datasets from PhishTank and UNB.

To compare the performance of all seven models (Decision Tree, Random Forest, Multi-layer Perceptrons, XGBoost, Support Vector Machine, Auto Encoder, Transformer), we randomly select 1000 new URLs other than the training and validation dataset. Half of these 1000 URLs are malicious URLs, and the other half are benign ones. All of these 1000 URLs are labeled. We used the 1000 URLs to conduct predictions with each one of the seven models. After the experimental prediction process, we compared the predicted results and the actual labels concerning each model. Each model's performance in the experiment was reflected by its corresponding test accuracy, precision, recall, and F1-score values. The result is in Table \ref{tab1}. According to the performance result table, we can find that our Transformer models got the highest values across all the four calculated metrics. For all these four metrics, the higher value suggests the better performance of the model from a specific aspect. We also sorted the models by the F1-score in descending order to reflect a model's all-around performance. The F1-score (\ref{eq8}) is the harmonic mean of precision (\ref{eq6}) and recall (\ref{eq7}). It takes the contribution of both, so the higher the F1 score, the better. A model does well in the F1 score if the positive predicted are positives (precision), does not miss out on positives, and predicts them negative (recall). Therefore, in terms of overall performance or performance from a specific aspect, our Transformer model presents the best performance among all seven models involved in the performance comparison experiment.

\begin{table}[htbp]
\begin{center}
\caption{Model Performance Comparison}
\resizebox{\columnwidth}{!}{
\begin{tabular}{|c|c|c|c|c|} 
\hline
\textbf{ML Model}       & \textbf{Test Accuracy} & \textbf{Test Precision} & \textbf{Test Recall} & \textbf{Test F1-Score}  \\ 
\hline
Transformer             & 0.973                  & 0.984                   & 0.962                & 0.973                   \\ 
\hline
Random Forest           & 0.849                  & 0.973                   & 0.718                & 0.826                   \\ 
\hline
Autoencoder             & 0.843                  & 0.934                   & 0.738                & 0.825                   \\ 
\hline
XGBoost                 & 0.825                  & 0.831                   & 0.816                & 0.823                   \\ 
\hline
Support Vector Machines & 0.845                  & 0.978                   & 0.706                & 0.820                   \\ 
\hline
Decision Tree           & 0.841                  & 0.975                   & 0.700                & 0.815                   \\ 
\hline
Multilayer Perceptrons  & 0.778                  & 0.747                   & 0.840                & 0.791                   \\
\hline
\end{tabular}
}
\label{tab1}
\end{center}
\end{table}

\section{Conclusion and Future Work}\label{sec:conclusion}
 In this paper, we reviewed literature about different approaching trends on malicious URL prediction. Based on the existing researches in this field, we introduced our construction of a Transformer classifier model for predicting the malicious URL. Additionally, we demonstrated our training dataset and the corresponding training processes. Also, we evaluated our finalized model by its performance upon the validation dataset and its quality compared to the other six machine learning or deep learning-based models. Comparing to the performance of the other six models (Decision Tree, Random Forest, Multi-layer Perceptrons, XGBoost, Support Vector Machine, and  Auto Encoder), we concluded that our transformer model is the best performing model from all perspectives among the total seven models when conducting predictions using our model comparison dataset. Future training based on the current finalized model with a large dataset may improve our robustness. The performance and robustness could be pushed onto a higher level, especially when using a huge dataset, including a large portion of short URLs. Even with the finalized version, our transformer model still provides an innovative idea of a low-cost but good-performing solution for malicious URL prediction. Also, our current findings and implementations of our transformer model verified that transformer technology is useful in malicious URL prediction and is worth future research and even productizations.

 \bibliographystyle{IEEEtran}
 \bibliography{IEEEabrv,cis6530_project_report.bib}




\end{document}